\documentclass[aip,prb,twocolumn,letterpaper,superscriptaddress,showpacs]{revtex4}
\usepackage{graphicx}
\usepackage{CJK}
\usepackage{amsmath,amssymb}

\usepackage{tikz}

\begin{document}
\begin{CJK*}{UTF8}{bsmi}
\title{
Effective reduction of the coercivity for Co$_{72}$Pt$_{28}$ thin film by exchange coupled Co$_{81}$Ir$_{19}$ soft layer with negative magnetocrystalline anisotropy
}
\author{J. Y. Jiao}
\author{T. Y. Ma}
\author{Z. W. Li}\email{zweili@lzu.edu.cn}
\author{L. Qiao}
\author{Y. Wang}
\author{T. Wang}\email{wtao@lzu.edu.cn}
\author{F. S. Li}

\affiliation{Institute of Applied Magnetics, Key Lab for Magnetism and Magnetic Materials of the Ministry of Education, Lanzhou University, Lanzhou 730000, P.R. China}
\date{\today}

\begin{abstract}
We report on the investigation of coercivity changes of the Co$_{72}$Pt$_{28}$/Co$_{81}$Ir$_{19}$ exchange-coupled composite (ECC) media with negative soft-layer (SL) magnetocrystalline anisotropy (MA) . Our results show that the hard-layer (HL) of our sample exhibits a columnar type microstructure with well isolated grains and the SL with hcp-structure grows on top of the HL with the same texture. Therefore, strong coupling of the two layers have been realized as evidenced by the magnetic characterization. Importantly, we observe a more effective reduction of the coercivity of the ECC media by using SLs with negative MA when compared to the use of SLs with positive or negligible MA. The experimental results are corroborated by theoretical calculations.
\end{abstract}


\maketitle
\end{CJK*}

\section{Introduction}
In order to solve the so called "magentic recording trilemma" problem, different approaches such as exchange-coupled composite (ECC) media \cite{apl.87.012504,apl.89.113105,pieee.96.1799,am.111.47} and heat/microwave/light assisted magnetic recording \cite{ieeeMT.49.718,apl.103.042413,prappl.6.054004}, have been used to reduce the coercivity of the recording media with very high uniaxial anisotropy constant. Among them,
the ECC media, in the form of magnetically hard/soft multi-layer structure, have attracted much attention due to their promising potential in resolving the above mentioned trilemma challenge \cite{apl.87.012504,apl.89.113105,pieee.96.1799,am.111.47}. Extensive theoretical work have been performed to investigate the effects of soft-layer (SL) thickness \cite{jmmm.290.551,apl.93.122502}, anisotropy constant \cite{apl.89.062512,jmmm.321.545,jmmm.320.3083}, hard-to-soft layer coupling strength \cite{pieee.96.1799,sr.5.16212,jap.99.08Q905,jmmm.316.159}, and the magnetization reversal process \cite{apl.91.182502,jap.99.08Q909,jap.99.08Q905}.
Experimentally, many work have been realized with FePt \cite{ieeeTM.46.1795,apl.95.022516,apl.96.032505,apl.100.142406,jap.105.123903,jap.111.103916}, CoPt \cite{apl.94.232502,jap.115.243905,jap.105.07B733,jmmm.323.2569,ieeeTM.43.2166} due to their high uniaxial magnetocrystalline anisotropy (MA) which is essential to maintain high thermal stability in high density magnetic recoding. Considerable reduction of the coercivity were achieved by changing the SL composition \cite{apl.100.142406,apl.94.232502,jap.115.243905} which alters its anisotropy constant, and by changing the SL and/or interlayer thickness \cite{ieeeTM.46.1795,apl.95.022516,apl.96.032505,jap.105.07B733,jmmm.323.2569,ieeeTM.43.2166} which optimizes the hard-to-soft layer coupling strength, thus improving the writability of the media while keeping high thermal stability.

However, all these experimental work were done with positive or negligible SL MA, $K_s$, in accordance with the theoretical work of Suess et al. \cite{jmmm.321.545}.
From the work of Suess et al., optimal reduction of the switching field can be realized with small but positive SL MA and the switching field will increase by increasing the anisotropy constant or by decreasing it towards the negative direction (see Fig. 2 in Ref. \cite{jmmm.321.545}).
On the other hand, using the model proposed by Victora et al. \cite{ieeeTM.41.537}, Wang et al. \cite{jmmm.320.3083} have shown that negative SL anisotropy actually has a beneficial contribution in decreasing the switching field.
To resolve the different theoretical predictions between Suess et al. and Victora et al. and to further optimize the magnetic properties of the ECC media, it is very interesting to experimentally investigate how does negative SL MA affect the coercivity of the ECC system.

Therefore, we report detailed studies on the reduction of its coercivity of the Co$_{72}$Pt$_{28}$/Co$_{100-x}$Ir$_{x}$ ECC system. By changing the Ir content of the Co$_{100-x}$Ir$_{x}$ SL, its MA can be tuned \cite{nrl.12.21,jap.99.08Q907}. Here, positive MA means that the easy axis of the magnetic material is along its crystallographic c-axis and negative MA, on the other hand, means that the easy axis is within its ab-plane. Thus, by oriented-growth of the hcp-Co$_{81}$Ir$_{19}$ (negative MA with $x=19$) layer with its crystallographic c-axis parallel to the normal of the film surface, its magnetic moments can be restricted strictly in-plane by its negative MA, thus providing a very strong negative total in-plane anisotropy for our ECC media \cite{sr.6.20140,jmmm.444.119}. A series of samples with different SL thickness and composition were investigated and we will show that negative SL MA is more efficient in reducing the coercivity of the ECC media, thus our results agree with the theoretical prediction of Victora et al. \cite{ieeeTM.41.537,jmmm.320.3083} and open a new playground for further improvements of the magnetic properties of the ECC media for high density magnetic recording.

\section{Experiments}

Co$_{72}$Pt$_{28}$/Co$_{81}$Ir$_{19}$ ECC media system ($K_s=-6\times10^5$\,J/m$^3$ \cite{nrl.12.21}) were prepared by DC perpendicular magnetron sputtering method with a layered structure of substrate/Ta(5\,nm)/Pt(10\,nm)/Ru(15\,nm)/Co$_{72}$Pt$_{28}$(20\,nm)
/Co$_{81}$Ir$_{19}$($t_{s}$\,nm) with $t_{s}$=0$\sim$15\,nm (see inset of Fig.\ref{Figtem}). For simplicity, our sample will be denoted as HL/SL19 where HL/SL stands for hard-layer/soft-layer and 19 represents the Ir content. Si(100)-orientation wafer with surface oxidation were used as substrate and the Ta(5\,nm)/Pt(10\,nm) layer was grown to promote Ru(002) texture which induces the columnar growth of the Co$_{72}$Pt$_{28}$ layer with well isolated grains \cite{jap.104.073904,ieeeIMC.2006.12}. For the growth of the HL and SL, Ar pressures of 4\,Pa and 0.3\,Pa were used, respectively. The base pressure before deposition was lower than 2$\times$10$^{-5}$\,Pa. For comparison reasons, ECC media systems with different SL compositions were also prepared with a similar fashion. To be exact, SL=Fe, Co and Co$_{100-x}$Ir$_{x}$ with x=23, 29 and 33 systems were prepared and these samples will be denoted as HL/(Fe,Co) and HL/SL(23,29,33), respectively. Note, that all Co$_{100-x}$Ir$_{x}$ with x=19, 23, 29 and 33 samples have the same crystal structure \cite{nrl.12.21,jap.99.08Q907}.

Chemical composition of our thin films have been characterized by energy dispersive spectrometer. Grain morphology of the sample was measured using a transmission electron microscope (TEM). The crystal structure of our sample was characterized by x-ray diffraction (XRD) with Cu $K_{\alpha1}$ radiation. Characterization of the magnetic properties were done with a vibrating sample magnetometer (VSM). Dynamic magnetic properties were measured with an electron spin resonance spectrometer (ESR) to determine the intrinsic magnetocrystalline anisotropy constant and the detailed procedure can be found in our previously published works \cite{sr.6.20140,jmmm.444.119}. These measurements were all done at room temperature and our samples were stored under vacuum when the measurements are done.

\section{Results and discussion}

\begin{figure}
\centering
\includegraphics[width=0.7\columnwidth,clip=true]{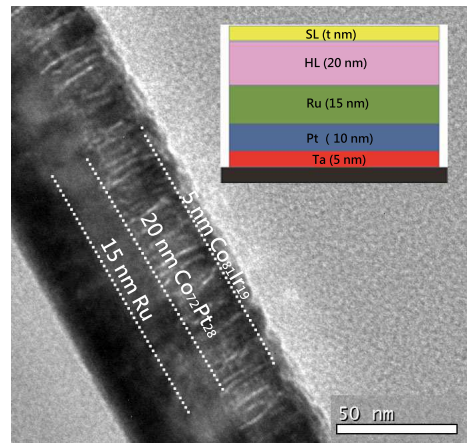}
\caption{
(color online)
TEM image of the cross section of the HL/SL19 sample with the SL thickness of 5\,nm. Inset shows the schematic layer structure.
}
\label{Figtem}
\end{figure}

\begin{figure}
\centering
\includegraphics[width=0.9\columnwidth,clip=true]{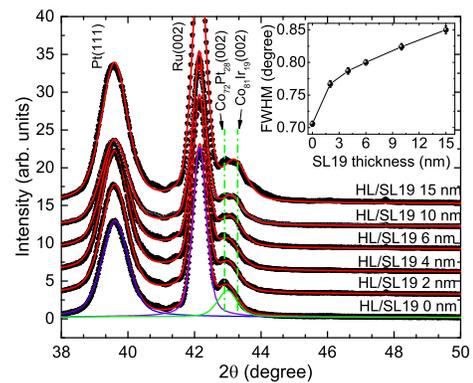}
\caption{
(color online)
XRD patterns of the HL/SL19 samples with indicated soft layer thicknesses. The solid lines represent profile fitting as described in the text. Inset shows the soft layer thickness dependence of the HL/SL19 (002) peak width.
}
\label{Figxrd}
\end{figure}

\begin{figure}
\centering
\includegraphics[width=0.9\columnwidth,clip=true]{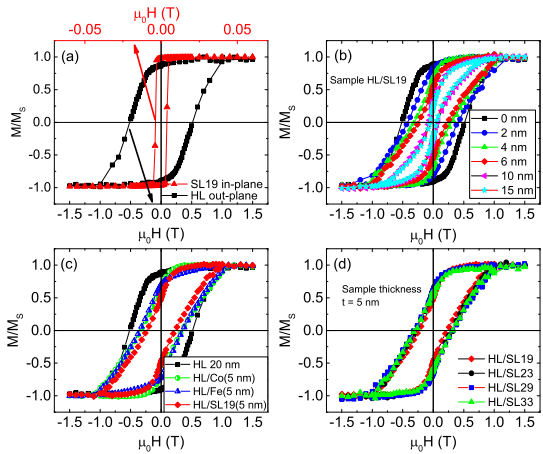}
\caption{
(color online)
(a) Magnetic hysteresis loops of the separate HL and SL samples grown on the same seed-layer with applied magnetic field along the easy magnetic direction of the sample indicating good hard/soft magnetic properties. Note the very different $\mu_0H$ scale. (b) Magnetic hysteresis loops of sample HL/SL19 with different SL thicknesses. (c) Comparison of the magnetic hysteresis loops between samples with no SL and Co(5\,nm), Fe(5\,nm) and SL19(5\,nm) as the SL. (d) Magnetic hysteresis loops of samples HL/SL(19-33) with various SL anisotropy constant. Note that the hysteresis loops in (c)-(d) have been measured with the external field applied perpendicular to the film plane.
}
\label{Figvsm}
\end{figure}

In Fig.\ref{Figtem} we present the typical TEM image of the cross section of sample HL/SL19 with $t_{s}$=5\,nm. Columnar growth of the Co$_{72}$Pt$_{28}$ layer with well isolated grains can be seen from the image and the well isolated nature of these grains are also reflected by the tilted magnetic hysteresis loop of the HL sample as shown later in Fig.\ref{Figvsm} (a) \cite{ieeeIMC.2006.12}.
The XRD pattern of our sample are shown in Fig.\ref{Figxrd}. All samples exhibit three main peaks at 39.6$^o$, 42.1$^o$, and 43.1$^o$ which correspond to Pt (111), Ru (002), and Co$_{72}$Pt$_{28}$/Co$_{81}$Ir$_{19}$ (002) peaks \cite{jap.104.073904,nrl.12.21}, respectively. The minor peak at 47.7$^o$ comes from the Si substrate. A closer check on the third peak at 43.1$^o$, one can find that the peak is actually composed of two peaks corresponding to Co$_{72}$Pt$_{28}$ and Co$_{81}$Ir$_{19}$ layers as denoted by the vertical green dash-doted lines in Fig.\ref{Figxrd}. However, trying to fit this peak with two sub-peaks was not successful due to their closeness and thus we fitted this peak with only one peak. The determined peak full width at half maximum (FWHM) was plotted as a function of the SL thickness, as shown in the inset of Fig.\ref{Figxrd}. One can see that FWHM increases with the SL thickness being consistent with the gradual increase in intensity of the SL sub-peak, inline with the gradual increase of the volume fraction of the SL. Importantly, the only observable (002) peak for the Co$_{72}$Pt$_{28}$ and Co$_{81}$Ir$_{19}$ layers indicate the oriented-growth of the these layers. Since Co$_{72}$Pt$_{28}$ (Co$_{81}$Ir$_{19}$) layer exhibit positive (negative) MA, the easy magnetic direction of the Co$_{72}$Pt$_{28}$ (Co$_{81}$Ir$_{19}$) layer is parallel (perpendicular) to the normal of the film surface \cite{jap.104.073904,nrl.12.21,sr.6.20140}.

Normalized magnetic hysteresis loops of our sample were shown in Fig.\ref{Figvsm} (a)-(d). In Fig.\ref{Figvsm} (a) we present the measurements for Co$_{72}$Pt$_{28}$ and Co$_{81}$Ir$_{19}$ separate layers grown on the same seed-layer. The measurements were done with the applied magnetic field along the easy magnetic direction of the HL (perpendicular to the film plane) or SL (parallel to the film plane) sample. For the HL sample, the easy-axis coercivity and saturation magnetization were determined to be 0.51\,T and 1.1$\times10^6$\,A/m, respectively. The MA of the HL was calculated to be 1.25$\times$10$^6$\,J/m$^3$ by the area difference between the hard-axis and easy-axis hysteresis loops. These magnetic parameters are close to reported values of similar composition \cite{jap.104.073904,jap.100.054909}. The tilt of the hysteresis loop suggests a weak inter-granular coupling of the HL, which is favorable for the application in perpendicular magnetic recording \cite{ieeeIMC.2006.12}. For the SL sample, an easy-axis coercivity of 0.004\,T and saturation magnetization of 1.2$\times10^6$\,A/m were obtained which are close to the values of our previous report \cite{nrl.12.21}. These results prove that, under the current film growth conditions, HL and SL films with good structural and magnetic properties can be obtained.

In Fig.\ref{Figvsm} (b), we present the measurements for our ECC sample HL/SL19 with different SL thicknesses, $t_{s}$=0$\sim$15\,nm. Clearly, the coercivity of the ECC system were greatly reduced. No obvious step were observed during the magnetization reversal process, indicating strong coupling between the hard and soft layers being consistent with the absence of any interlayers between the two layers, which shows the uniform switching of the SL with respect to the HL. However, as the SL thickness $t_{s}$ increases, the perpendicular magnetic anisotropy of the ECC media is significantly degraded as the rectangular loops tend to tilt more and more with increasing $t_{s}$. This corresponds to a reduction of the squareness ratio of the hysteresis loops, which indicates the domination of the negative anisotropy of the SL, resulting in a reduction of the remanence of hysteresis loops \cite{jap.115.243905}.

For comparison, we also prepared HL/Co(5\,nm) and HL/Fe(5\,nm) ECC samples with small but positive SL MA and their magnetic hysteresis loops are shown together with that of HL/SL19(5\,nm) in Fig.\ref{Figvsm} (c). Obviously, the SL19 layer with negative MA is more efficient in reducing the coercivity of the ECC system. Additionally, to study how the reduction of the coercivity depends on the strength of the anisotropy constant of the SL for the ECC media, we prepared different samples with different Ir content in the SL, thus different SL MA \cite{nrl.12.21,jap.99.08Q907}, as shown in Fig.\ref{Figvsm} (d). One can see that the strength of the anisotropy constant does not have significant impacts on the media coercivity as the SL thickness does.

\begin{figure}
\centering
\includegraphics[width=0.9\columnwidth,clip=true]{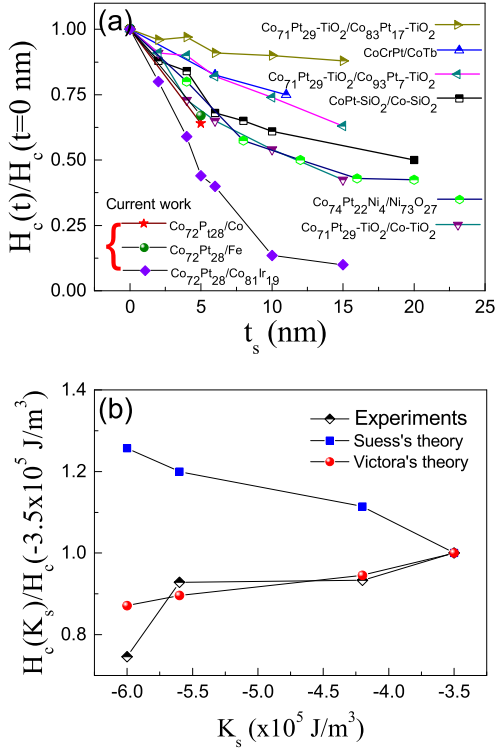}
\caption{
(color online)
(a) Reduction of the coercivity of the ECC media for various HL/SL systems as a function of the SL thickness. For comparison reasons, the data were normalized to the initial value with zero SL thickness. Data with HL composition of Co$_{71}$Pt$_{29}$-TiO$_2$ was taken from Ref.\cite{jmmm.323.2569}, CoPt-SiO$_2$ was taken from Ref.\cite{jap.105.07B733}, CoCrPt was taken from Ref.\cite{jap.91.8058}, and Co$_{74}$Pt$_{22}$Ni$_4$ was taken from Ref.\cite{ieeeTM.43.2166}. (b) Influence of the strength of the SL MA on the coercivity of the ECC media with a SL thickness of 5\,nm. Theoretical calculations were done using equation \ref{eqSuess} (Suess's theory) and equation \ref{eqWang} (Victora's theory) for comparison as described in the main text.
}
\label{Figres}
\end{figure}

To see the coercivity reduction more clearly, we summarized our results in Table \ref{table} and Fig.\ref{Figres} together with published data on the CoPt system with different SLs. Obviously, as shown in Fig.\ref{Figres} (a), Co$_{81}$Ir$_{19}$ SL with negative MA is more effective in reducing the ECC media coercivity for any defined SL thickness. In Fig.\ref{Figres} (b), we present the reduction of the media coercivity as a function of the SL MA constant. The data have been normalized to the value of the most right side data point at $K_s$=-3.5$\times$10$^5$\,J/m$^3$ for better comparison with calculations. The enhanced reduction with decreasing $K_s$ is inconsistent with the predictions by Suess et al. \cite{jmmm.321.545}; however being in agreement with the results of Victora et al. \cite{ieeeTM.41.537,jmmm.320.3083}.

\begin{table}[hbt]
\centering
\caption{Published data on the coercivity $\mu_0H_C$ reduction for the Co$_{100-x}$Pt$_x$ system using different soft layers. t$_{s}$ denotes the hard/soft-layer thickness and * indicates the current work.}
\label{table} {
\begin{tabular}{c c c c}
\hline \hline
Hard/soft-layer   &  t$_{s}$ (nm) & $\mu_0H_C$ (T) & ref.       \\
\hline
CoPt-SiO$_2$/Co-SiO$_2$ & 20/0-20 & 0.7-0.35 & \cite{jap.105.07B733} \\
CoCrPt/CoTb & 20/0-11 & 0.4-0.3 & \cite{jap.91.8058} \\
Co$_{71}$Pt$_{29}$-TiO$_2$/Co-TiO$_2$ & 15/0-15 & 0.47-0.2 & \cite{jmmm.323.2569} \\
Co$_{71}$Pt$_{29}$-TiO$_2$/Co$_{93}$Pt$_{7}$-TiO$_2$ & 15/0-15 & 0.47-0.3  & \cite{jmmm.323.2569} \\
Co$_{71}$Pt$_{29}$-TiO$_2$/Co$_{83}$Pt$_{17}$-TiO$_2$ & 15/0-15 & 0.47-0.42  & \cite{jmmm.323.2569} \\
Co$_{74}$Pt$_{22}$Ni$_4$/Ni$_{73}$O$_{27}$ & 12/0-20 & 0.75-0.3  & \cite{ieeeTM.43.2166} \\
Co$_{72}$Pt$_{28}$/Co$_{81}$Ir$_{19}$ & 20/0-15 & 0.52-0.052  &  * \\
\hline \hline
\end{tabular}}
\end{table}

\begin{figure}
\centering
\includegraphics[width=0.9\columnwidth,clip=true]{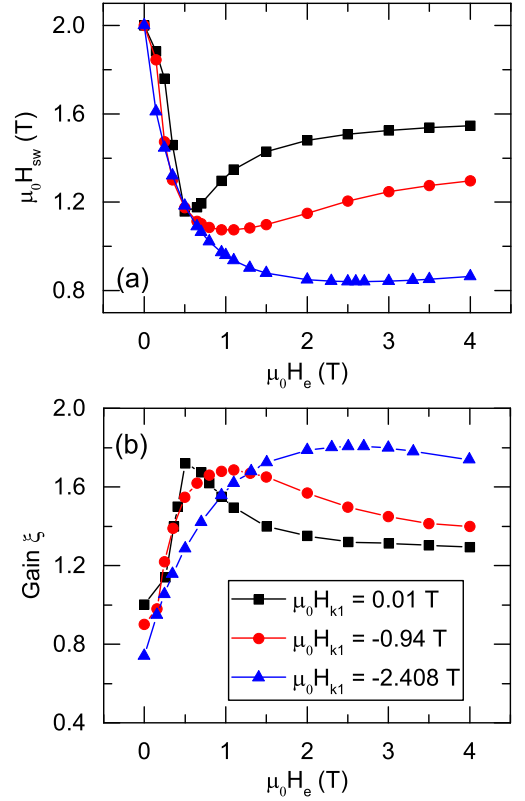}
\caption{
(color online)
(a) Switching field H$_{sw}$ as a function of the interlayer exchange field H$_e$ with various soft layer anisotropy H$_{k1}$. (b) Dependence of the gain factor $\xi$ on H$_e$ with various H$_{k1}$.
}
\label{Figcalc}
\end{figure}

To understand this results more clearly, we would like to revisit the theoretical model used by Suess et al. and Wang et al.. In the work of Suess et al. \cite{jmmm.321.545}, they calculated the switching field with a simple expression $H_s=max(H_p, H_n)$ with
\begin{eqnarray}
H_n=\frac{2K_{s}}{\mu_0M_1}+\frac{2A\pi^2}{4t_s^2\mu_0M_1} \nonumber\\
H_p=\frac{1}{4}\times\frac{2(K_{h}-K_{s})}{\mu_0M_{2}},
 \label{eqSuess}
\end{eqnarray}
where $\mu_0H_n$ and $\mu_0H_p$ represent the nucleation field in the soft layer and pinning field at the interface, respectively. $K_s$ and $K_h$ denote the anisotropy constants of the soft and hard layers. $A$ is the intragranular exchange constant and $M_1$/$M_2$ is the saturation magnetization of the soft/hard layer. $t_s$ is the soft layer thickness. On the other hand, Victora et al. \cite{ieeeTM.41.537,jmmm.320.3083} used a more complete model by minimizing the total energy of the ECC system.
In this model, the total magnetic energy of the ECC system can be expressed by
\begin{eqnarray}
 E = HM_1 cos\theta_1V_1+HM_2cos\theta_2V_2+K_1sin^2\theta_1V_1  \nonumber\\
+K_2sin^2\theta_2V_2-J_ecos(\theta_1-\theta_2),
 \label{eqWang}
\end{eqnarray}
where $\mu_0H$, $M_i$, $K_i$, and $V_i$ denote the external field, saturation magnetization, anisotropy constant and volume of the particle, respectively. Sub-index 1, 2 indicate the soft, hard region, respectively. $J_e$ is the exchange constant and $\theta_i$ indicates the angle between the magnetic moments and the external filed.
In addition, demagnetizing energy was included properly in this model which was omitted by Suess et al. in their calculation \cite{jmmm.321.545}.

In the calculation work of Wang et al., Victora's model was used and the negative SL anisotropy was assumed to come from the demagnetizing field which is limited by the largest demagnetizing factor ($N=1$) by setting the MA to zero \cite{jmmm.320.3083}. Here, we further push the total SL anisotropy towards the negative direction by using Co$_{1-x}$Ir$_{x}$ with negative MA (total in-plane anisotropy field can be expressed as $H_{k1}=-M_s+2K_s/\mu_0M_s$). As shown in Fig.\ref{Figcalc}, we recalculated the switching field $\mu_0H_{sw}$ and Gain factor $\xi$ as a function of the interlayer exchange field $\mu_0H_{e}$ using the same procedure as done by Wang et al \cite{jmmm.320.3083} with the total in-plane anisotropy field $\mu_0H_{k1}$ up to a negative value of $-2.408$\,T. The Gain factor, which is a reflection of the thermal stability, is defined as $\xi=2\Delta E/(H_{sw}M_aV)$, where $\Delta E$, $M_a$ and $V$ are the total anisotropy energy, average saturation magnetization, and total volume, respectively.
From Fig. \ref{Figcalc} (a)-(b), one can see that in the strong coupling regime (bigger $\mu_0H_{e}$), the more negative of the SL MA the more efficient in reducing the switching field and the bigger of the Gain factor can be obtained.
According to these results, the same conclusion can be drawn as in reference \cite{jmmm.320.3083}, that is by using a SL with negative $K_s$ can further reduce the switching field of the ECC system and can further increase the Gain factor $\xi$ even up to very large negative values of the MA ($K_s$=-6$\times$10$^5$\,J/m$^3$ in our case).

In order to reproduce the same trend of our experimental results, we calculated the SL MA dependence of the coercivity reduction using the above two theories as shown in Fig.\ref{Figres} (b). Clearly, in opposite to the results of Suess et al., Victora's theory predicts the same trend with our experiments. For the calculation, $A=10^{-11}$\,J/m$^3$ and exchange field of $\mu_0H_e=4$\,T were used. The SL thickness was set to be 5\,nm which is the same as our experiments, the anisotropy constant and the saturation magnetization of the HL and SL were taken from the measured values in this work. First, calculation of the switching field were done separately using equation (\ref{eqSuess}) \cite{jmmm.321.545} and equation (\ref{eqWang}) \cite{jmmm.320.3083}. Then, they were normalized together with experimental data at the most right point ($K_s$=-3.5$\times$10$^5$\,J/m$^3$) for comparison of the trends with changing MA. As shown by equation (\ref{eqWang}), Victora's model starts from a more general way and the switching field can be calculated by minimizing the total energy against $\theta_1$ and $\theta_2$ \cite{ieeeTM.41.537,jmmm.320.3083}. The magnetostatic field coming from other grains of the media is considered as external field. Most importantly, it points out that the demagnetizing energy is not a negligible factor for the ECC media since it has an obvious contribution to the anisotropy of the SL. We further notice that Suess's model predicts an decrease of the thermal stability of the recording media with negative magnetocrystalline anisotropy of the SL \cite{jmmm.321.545}. On the other hand, in our calculation using Victora's model, we can get enhanced thermal stability for stronger inter-layer coupling regimes (larger Gain factor in our calculation as discussed above.).

\begin{figure}
\centering
\includegraphics[width=0.9\columnwidth,clip=true]{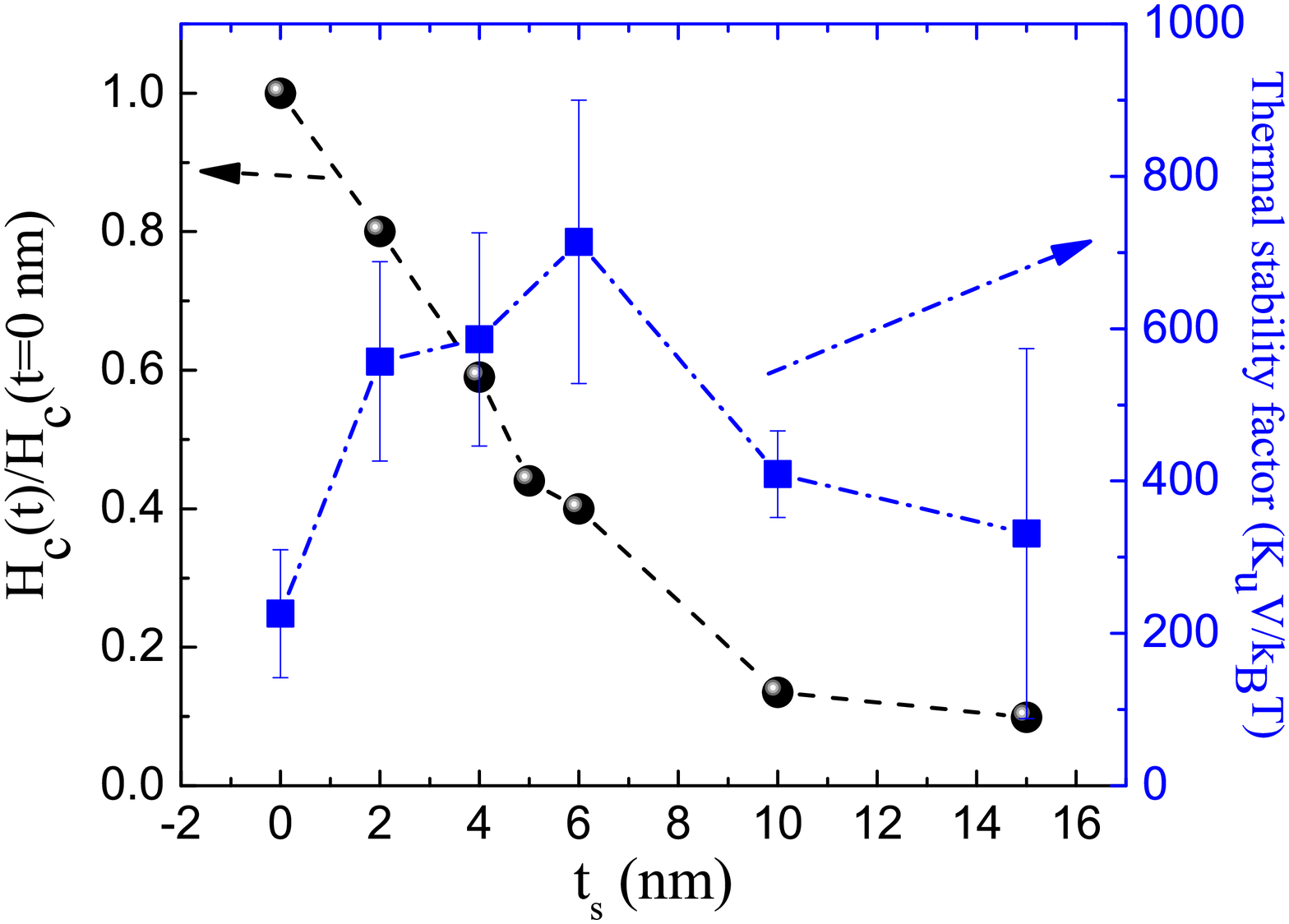}
\caption{
(color online)
Thickness dependence of the normalized coercivity (left axis) and thermal stability factor (right axis) of our sample HL/SL19 .
}
\label{FigTSF}
\end{figure}

To check the thermal stability of our ECC media, we measured the remnant coercivity $\mu_0H_r$ as a function of waiting time to study the magnetic decay of our sample. First, a negative saturation field of 2\,T was applied, then the field was increased to positive fields around the coercive field for some waiting time $t$, after that the field was set to zero and measure the magnetic moment of the film. The remnant coercivity $\mu_0H_r(t)$ can be obtained by linear fitting of the measured M(H) curve for each waiting time $t$. The thermal stability factor $K_uV/k_BT$ can then be evaluated by analyzing the time dependent $\mu_0H_r(t)$ using Sharrock's equation \cite{jap.76.6413},
\begin{eqnarray}
Hr(t)=H_{r}(0)\{1-[\frac{k_BT}{K_uV}\ln(\frac{f_0t}{\ln2})]^{1/n}\},
 \label{eqSharrock}
\end{eqnarray}
where $f_0$ is called the attempt frequency ($\sim10^{10}$\,Hz for magnetic systems \cite{jap.76.6413}), n=1.5 was fixed \cite{prb.75.174430}. $K_uV$ is the energy barrier at zero applied field, and $k_BT$ is the thermal energy required to flip the magnetization. The obtained $K_uV/k_BT$ are shown in Fig. \ref{FigTSF} together with the coercivity reduction of our HL/SL19 sample.
As can be seen, different with the monotonic decrease behavior of the coercivity, the thermal stability factor first increase with the SL thickness up to 6\,nm and then decrease with further increase of the SL thickness. Similar to the FePt-C/FePt \cite{am.111.47} system, the initial increase can be attributed to the increase of the switching volume due to the strong exchange coupling between the HL and SL, and the decrease above 6\,nm is mainly due to the demagnetizing effect. Typical $K_uV/k_BT$ for the current perpendicular recording media using CoCrPt system is around 80 \cite{jap.102.011301}, which is much smaller than our value of 226 at zero SL thickness. Therefore, our ECC media show very good thermal stability and writability than the current CoCrPt perpendicular media.
Finally, we would like to emphasize the advantage of Victora's model in the study of ECC media systems and the rather regretful fact that no experimental work using SLs with negative MA have been employed to reduce the coercivity of the ECC media. We hope that our work will promote further experimental work in this direction and promote further improvements of the writability for high density magnetic recording.

\section{Summary}

In conclusion, we have prepared Co$_{72}$Pt$_{28}$/Co$_{81}$Ir$_{19}$ ECC media with negative SL MA. TEM image shows the columnar growth of the Co$_{72}$Pt$_{28}$ HL and XRD measurements shows the oriented-growth of the SL on top of the HL with hcp-structure. Good hard/soft magnetic properties of the hard/soft layers have been proved by our VSM measurements. We observe a single step magnetization reversal of the ECC media, suggesting strong coupling of the hard and soft magnetic layers. More interestingly, we observe an enhanced reduction of the coercivity of the ECC media by using SLs with negative MA when compared to the use of SLs with small but positive MA. These results have been qualitatively reproduced by theoretical calculations. Therefore, our results provide a new direction for us to further improve the writability of the ECC media for future high density magnetic recording.

\section{Acknowledgement}

This work was supported by the National Natural Science Foundations of China (Nos. 11574122 and 11204115) and the Fundamental Research Funds for the Central Universities (Nos. lzujbky-2017-k20 and lzujbky-2017-31).

\section*{References}
\bibliography{CoPtCoIr_ref}

\end{document}